\documentstyle[preprint,aps,eqsecnum,epsfig]{revtex}
\begin{document}

\draft

\preprint{\fbox{\sc version of \today}}

\def\lt{\raisebox{0.2ex}{$<$}}
\def\gt{\raisebox{0.2ex}{$>$}}

\title{Global Properties of Spherical Nuclei Obtained from
Hartree-Fock-Bogoliubov Calculations with the Gogny Force\thanks{The work was
partially sponsored by the Polish Committee of Scientific Research KBN
No.~2P~03B~115~19, POLONIUM No.~017~04~UG (2001) and IN2P3 Convention 99-95.}}

\author{            M. Kleban, B. Nerlo-Pomorska\\
{\it Katedra Fizyki Teoretycznej, Uniwersytet Marii Curie Sk\l odowskiej,}\\
                   {\it PL-20031 Lublin, Poland }\\
         J. F. Berger, J. Decharg\'e, M. Girod, S. Hilaire \\
         {\it Centre d'Etudes de Bruy\`eres-le-Ch\^atel,
              F-91680 Bruy\`eres-le-Ch\^atel, France}}


\maketitle

\begin{abstract}
Selfconsistent Hartree-Fock-Bogoliubov (HFB) calculations
have been performed with the Gogny force
for nuclei along several constant $Z$ and constant $N$ chains,
with the purpose of extracting the macroscopic
part of the binding energy using the Strutinsky prescription. The
macroscopic energy obtained in this way is compared to current liquid
drop formulas.
The evolution of the single particle levels derived from the HFB calculations
along the constant $Z$ and constant $N$ chains
and the variations of the different kinds of nuclear radii are also analysed.
Those radii are shown to follow isospin-dependent three parameter laws
close to the phenomenological formulas which reproduce
experimental data.
\end{abstract}


%
%
%
%

\section{Introduction}
\label{Sect1}
Microscopic calculations using the effective nucleon-nucleon interaction
proposed by Gogny
have been shown to reproduce
in very satisfactory way  a variety of nuclear properties
over a wide range of proton and neutron numbers \cite{dech}.
For instance,
binding energies, separation energies of one or two nucleons and
charge and neutron mean square radii usually agree quite well with known
experimental data.
In view of this,  it seems appropriate to use
theoretical results obtained in such a framework
to derive systematics of nuclear properties that could be useful for
making predictions concerning nuclear species or
aspects of nuclear structure no yet known as neutron distribution radii or
binding energies of superheavy nuclei.

In the present work, simple formulas are used in order to represent
the binding energies and the charge, proton and neutron root-mean-square
(r.m.s.) radii obtained in HFB calculations with the Gogny force, and
they are compared with existing phenomenological formulas.
Our aim is to determine simple but realistic enough parametrizations
that would avoid performing time-consuming microscopic calculations
when only rough estimates of the binding energies or radii are needed.
Some preliminary results have already been published in \cite{pom,kleb},
for a  smaller number of nuclei however.

Section~\ref{Sect2} presents a short overview of the theoretical models used
in this work.
Section~\ref{Sect3}
 explains how the basis parameters employed in the self-consistent
calculations have been determined.
In Section~\ref{Sect4}, the results concerning the single
particle levels schemes of representative isotonic and isotopic chains are
analysed.
Section~\ref{Sect5} gives the method used to substract the
neutron and proton shell and pairing corrections
from the HFB energy.
The "macroscopic Gogny energy" obtained in this way
is then parametrized by means of a liquid-drop like formula~\cite{mye}.
In Section~\ref{Sect6}, systematics of the mass, charge, proton
and neutron  radii and of the ratios of proton to neutron radii
are presented.
Simple isospin- and mass number-dependent parametrizations
of these quantities are determined, which appear in excellent
agreement with those derived using the
relativistic mean field method \cite{rin}.
The formula obtained for the deformation-independent
ratios of proton to neutron
r.m.s. radii is especially useful, since it should be
valid for all even-even nuclei, either spherical or deformed.
Conclusions and perspectives for further work are gathered
at the end of the paper.

\section{Theory}
\label{Sect2}
The Gogny density-dependent effective nucleon-nucleon force  is of the
following form~\cite{dech}:
\begin{eqnarray}
V_{12}& = & \sum_{i=1}^2  \ (W_i + B_i\hat P_\sigma - H_i\hat P_\tau
-M_i\hat P_{\sigma}\hat P_{\tau}) \ e^{-\frac {\displaystyle
(\vec r_1- \vec r_2)^{2}} {\displaystyle\mu_i^{2}}}
\nonumber \\
& + & i \ W_{\rm LS}{\displaystyle \
(\overleftarrow{\nabla_1 - \nabla_2})\times\delta(\vec r_1 - \vec
r_2)(\overrightarrow{\nabla_1 - \nabla_2})} \cdot
 (\vec \sigma_1 + \vec \sigma_2)
\label{force} \\
& + & t_0 \ (1+x_0 \hat P_\sigma) \
\delta{\displaystyle (\vec r_1 - \vec r_2)}
\left[\rho(\frac{\displaystyle  \vec r_1 + \vec r_2 }
                {\displaystyle 2}) \right]^\gamma
+ V_{\rm Coul} \,\,, \nonumber
\end{eqnarray}
which represents a central finite range interaction, a
zero-range spin-orbit term and a zero-range density dependent
interaction, respectively,  to which one has to add the Coulomb interaction
in the case of protons.
The central interaction is made up of two distinct gaussians
with ranges $\mu_1$ and $\mu_2$ whose values are given below.
$\hat P\sigma$ and $\hat P\tau$ denote the
spin and isospin exchange operators respectively, and $\rho$ represents
the total density. \\
We use the Gogny  D1S \cite{ber} interaction the parameters of which are given
below:
\begin{eqnarray}
W_1 = -1720.30 \;\ {\rm MeV} & \, \, \;\   W_2 = \,\, 103.639 \;
\ {\rm MeV} \nonumber  \\
B_1 =\;\ \, 1300.00 \;\ {\rm MeV} & \;\  \;\  B_2 = -163.483 \;\ {\rm MeV}
\nonumber \\
H_1 = -1813.53 \;\ {\rm MeV} & \, \, \, \;\  H_2 = \, \, \, \, \,  162.812 \;
\ {\rm MeV}   \nonumber \\
M_1=\;\ \, 1397.60 \;\ {\rm MeV} & \, \, \,   \;\ M_2=-223.934 \;\ {\rm MeV}
\label{D1S} \\
\mu_1= \;\, \qquad 0.7 \,\;\ \;\ \, {\rm fm} &  \;\   \mu_2= \qquad   1.2 \;
\,\,\,\,  {\rm fm} \nonumber \\
t_0=1390.6 \;\ {\rm MeV}\, {\rm fm}^{3(1+\gamma)} &  \! \! \! \! \! \! \! \! \!
 \! \!  \! \! \! \!   x_0= \qquad 1 \nonumber \\
 \gamma= \qquad 1/3 \quad \ \qquad  & \ \ \ W_{LS}= \quad 130 \ {\rm MeV} \,
 {\rm fm}^5 \nonumber
\end{eqnarray}
When pairing correlations are neglected, the HFB approach reduces to
the Hartree-Fock (HF) method which determines a self-consistent
approximation of the nuclear mean-field.
The corresponding ground state energy $E_{HF}$ includes a contribution
$E_{\rm shell}$ from  shell effects which can be evaluated by applying
the Strutinsky smearing procedure \cite{strut}
to the HF single-particle level distribution.
The remaining part of $E_{HF}$  can be considered as a macroscopic, liquid-
drop like contribution $\tilde E$.
Denoting by  $e_{\nu}$ the HF proton or neutron single-particle energies,
the Strutinsky shell correction energy $E_{\rm shell}$ is:
\begin{equation}
  E_{\rm shell} = \sum_\nu 2 e_\nu - \tilde E \,,
\label{shell}
\end{equation}
where $\tilde E$ is the smoothed energy
\begin{equation}
\tilde E=2\int_{-\infty}^{\lambda} e \bar{\rho}(e) de \,.
\label{smooth}
\end{equation}
With the Strutinsky prescription,
the smoothed density $\bar{\rho}$ is given by
\begin{equation}
\bar{\rho}(e)=\frac{1}{\gamma \sqrt{\pi}} \int_{-\infty}^{+\infty} \rho(e')\,
e^{\displaystyle-(\frac{e-e'}{\gamma})^2}f(\frac{e-e'}{\gamma})\,de' \,,
\label{rho}
\end{equation}
where $\lambda$ is the Fermi energy, $f$ the Strutinsky
(6$^{\rm th}$ order) curvature correction polynomial and  $\gamma$ the width
over which smoothing of the single particle level scheme is performed.
The reliability of this procedure requires that $E_{\rm shell}$
displays a plateau when drawn as a function of 
$\gamma$, separately for protons and neutrons.
A crucial parameter in this respect is
the number of single particle levels taken into account in the evaluation of
the  shell correction. This problem will be discussed in
Section~\ref{Sect5}.

Subtracting from the Hartree-Fock energy the shell correction energies
of neutrons $E_{\rm shell}^n$ and of protons $E_{\rm shell}^p$, one gets
the following estimate of the macroscopic part of the binding energy:
\begin{equation}
 E_{\rm macr} = E_{HF} - E_{\rm shell}^n - E_{\rm shell}^p \,.
\label{macr}
\end{equation}

Another correction to the macroscopic part of the binding energy is the
contribution $E_{\rm pair}$ of pairing correlations. This correction
has been calculated for each nucleus by taking the
difference between the energy $E_{HFB}$ obtained in a full HFB
calculation with the Gogny force and the energy $E_{HF}$ computed using
the simple HF method:
\begin{equation}
 E_{\rm pair} = E_{HFB} - E_{HF} \,.
\label{pair}
\end{equation}

The obtained r.m.s. radii of the neutron, proton, charge and mass
density distributions are obtained in the microscopic HFB calculations in
the usual way.
In deformed nuclei, these different radii contain the effect of
the quadrupole and other deformations \cite{nerl}.
However, the ratios of proton to neutron radii
are almost deformation-independent since
the proton and neutron density distributions are very similarly deformed.

Starting from the HFB mean square radii $\langle r^{2}\rangle$,
equivalent spherical liquid-drop radii $R$ can be defined through the
expression:
\begin{equation}
R=\sqrt{\frac{5}{3}}\sqrt{\langle r^{2} \rangle}\,.
\label{ldr}
\end{equation}
which follows from the formula giving the
mean square radius of a uniform spherical density distribution of radius $R$:
\begin{equation}
\langle r^{2}\rangle= {3\over 5} R^2 \,.
\label{ever}
\end{equation}

These equivalent radii have then been fitted to a three parameter formula,
with explicit isospin and  $A$ dependence 
\begin{equation}
R=r_{00}(1+\alpha \frac{N-Z}{A} +
\frac{\beta}{A})A^{\frac{1}{3}}=r_{0}(A,I)A^{\frac{1}{3}}\, .
\label{rphen}
\end{equation}
similar to the formula used in Ref.~\cite{war} in the analysis of the radii
obtained from  relativistic mean field calculations.
As the ratios of the proton and neutron r.m.s. radii are
almost deformation independent, they could be well approximated by
the following formula depending
only on the numbers of protons and neutrons:
\begin{equation}
\frac{r_{p}}{r_{n}}= c (1+ a\frac{N-Z}{A} + \frac{b}{A})\,.
\label{ror}
\end{equation}
This last relation, together with the effect of the non-point like charge
distribution of the proton evaluated with the
approximate formula:
\begin{equation}
\langle r_{ch}^{2}\rangle=\langle r_{p}^{2}\rangle+0.64\  {\rm fm}^{2}\, ,
\label{averch}
\end{equation}
can be useful for estimating the neutron radius of a
nucleus when its r.m.s. charge radius
$r_{ch} = \sqrt{\langle r_{ch}^{2}\rangle}$ is known:
\begin{equation}
r_{n}=\frac{\sqrt{\langle r_{ch}^{2}\rangle  - 0.64 {\rm fm}^2}}{c (1 +
       a\frac{N-Z}{A} + \frac{b}{A})}\, .
\label{rn}
\end{equation}

\section{Parameters of the calculation}
\label{Sect3}

The nuclei studied in the present work
are represented by crosses and dots in the  ($N$,$Z$) plot of Fig.~\ref{fig1}.
These are the nuclei close to magic proton and neutron numbers
and along the $\beta$-stability line whose ground states are expected to be
spherical \cite{mol}. They include Ca, Sr, Sn, Sm, Pb
and  Th isotopes,
the $N$=50, 82 and 126  isotone chains between the proton and neutron drip
lines and a few $\beta$-stable spherical nuclei
between $A=38$ and $A=218$.

In the microscopic HF and HFB calculations, the self-consistent equations
have been solved in matrix form by expanding  the single particle
or quasiparticle states on finite bases made of spherical
harmonic oscillator  (HO) eigenfunctions. These bases
depend on two parameters: the number
$N_o$ of major HO shells included in the bases and
the oscillator parameter $\hbar\omega$.

In the present study, bases with $N_o$=12, 14, 16 or 18 major shells
have been taken into account, depending on the nucleus under study, the criterium being
that $N_o$ is large enough to ensure convergence of the HFB energy ($E_{\rm HFB}$)
within a few keV.
For each nucleus and each value of $N_o$, the parameter $\hbar\omega$
has been chosen as the value
$\hbar\omega_{min}$ that minimizes the  HF energy $E_{\rm HF}$.
The values of $\hbar\omega_{min}$ obtained for Ca, Sn and Pb isotopes, and
for the $N$=126 isotones
are plotted in Fig.~\ref{fig2} as a function of $A$
for $N_o$=14, a value of $N_o$ large enough for all these nuclei.
The error bars indicate
the ranges of $\hbar\omega$ for which the variation of $E_{HF}$ does not
exceed 100 keV. One can see that,
with this number of HO shells,
relatively large changes in $\hbar\omega$ do not significantly affect
calculated values of $E_{HF}$. Consequently, approximate analytical
interpolation formulas can be used.

A large scale investigation of the values found in spherical nuclei
shows that an interpolation formula such as:
\begin{equation}
\hbar \omega_{min}=n (1+k \frac{N-Z}{A})A^{-\frac{1}{3}}
\label{hw1}
\end{equation}
can be adopted for all even-even nuclei, with
n and k depending on the number $N_o$ of shells included in the bases.
In the present work,
all the obtained values of $\hbar\omega_{min}$ could be fitted with the
values
$n$=64.05 MeV and $k$=-0.46.
A even more accurate interpolation formula, represented by the solid curve in
Fig.~\ref{fig2}, has been obtained for the specific set of
nuclei for which $N_{o}$=14 is found to be a sufficiently large number:
\begin{equation}
\hbar\omega_{min}=(0.0002A^{2} - 0.1A + 21.1)\ {\rm MeV}
\label{hw2}
\end{equation}
This formula gives a better agreement to calculated  $\hbar \omega_{min}$
for the Ca up to the Pb isotopes and for the $N$=126 isotones.

\section{Single particle levels}
\label{Sect4}
In order to better visualize the origin of shell effects in the
nuclei considered in this work,
the Hartree-Fock single-particle energies obtained with the Gogny force
are plotted in Figs.~\ref{fig3a}, \ref{fig3b} and~\ref{fig3c}
for neutrons (left hand sides) and protons (right hand sides).
The three figures correspond to $N$=50,  $N$=82 and  $N$=126  isotones,
respectively, and the levels are drawn as functions of the  proton number $Z$.
They are
labelled with the usual radial quantum numbers $n$, orbital
angular momentum $l$ and total angular momentum $j$. The
numbers $N_S$ between the levels indicate the number of nucleons necessary
to fill all the levels located below.

One can see that,
in addition to the magic shells corresponding to $N_S$= 20, 28, 50, 82, 126
and 184, well-marked subshells appear depending on the 
number of neutrons $N$.
For instance, subshells with  $N_S$=40 are clearly visible in the
neutron and proton level schemes of the $N$=50 and $N$=82 isotones, and
$N_S$=64 also appears as a subshell in the  $N$=126 isotones.

In Figs.~\ref{fig4a} to \ref{fig4f} the neutron (left hand sides) and proton
(right hand sides) single particle levels of the Ca, Sr, Sn, Sm, Pb and Th
isotopes are plotted as functions of the neutron number $N$.
Subshells corresponding to $N_S$=40, 64 or 100 can also be
observed in these figures besides the usual magic numbers
$N_S$=50, 82, 126, 184.
It is interesting to note that the magnitude of the proton and
neutron $N_S$=50 shell gaps steadily decreases from $\simeq$ 6 MeV down to
$\simeq$ 4 MeV as the proton number $Z$
increases from 20 to 90. This decrease does not occur in the case of the
other magic numbers $N_S$=82 and $N_S$=126.

\section{Microscopic and macroscopic part of binding energies}
\label{Sect5}

As explained in Section \ref{Sect2}, the HF single-particle levels
shown in Figs.~\ref{fig3a} to \ref{fig4f} have been used to separate
HF binding energies into two components by means of Strutinsky's method:
a macroscopic part behaving smoothly as a function of $A$ and a
shell correction reflecting the neutron and proton shells.
The method of Strutinsky is a subtle technique
which demands a very careful choice of the number of
single-particle states above the Fermi level included in the smoothing
procedure so as to obtain a plateau in the variations of the shell corrections with respect
to the smoothing width $\gamma$ .
In particular, including too many single particle states with positive
energies strongly affects the determination of the shell effects associated
with occupied orbitals.
This is so because single particle states with positive
energies are obtained in the present microscopic approach as a discrete set
whose energy spectrum and shell structure can strongly depend on the choice
of basis parameters.

For this reason, the HF single-particle proton and
neutron levels included in the Strutinsky smoothing procedure have
been chosen in the following way:
For lighter nuclei, a cut off in the single particle energy
has been  introduced, having the usual value $\lambda+2\hbar\omega_{0}$,
with $\lambda$ the proton or neutron Fermi energy and $\hbar\omega_{0}$
the average spacing between major shells. The number of
single-particle levels taken into account in this way is of the order of 30,
with maximum energies 5 MeV for neutrons and 15 MeV for protons above the Fermi
level.
In heavier nuclei, the number of levels included in the smoothing
procedure was increased to 50, which
corresponds to a cut-off in the single particle energies of the order of
$\lambda+4\hbar\omega_{0}$, {\it i.e.}
15 MeV for neutrons and 30MeV for protons.
These numbers have been found to provide a satisfactory stabilization of
shell corrections in all the nuclei studied in this work.

\begin{figure}

  \begin{center}

  \leavevmode

\epsfig{file=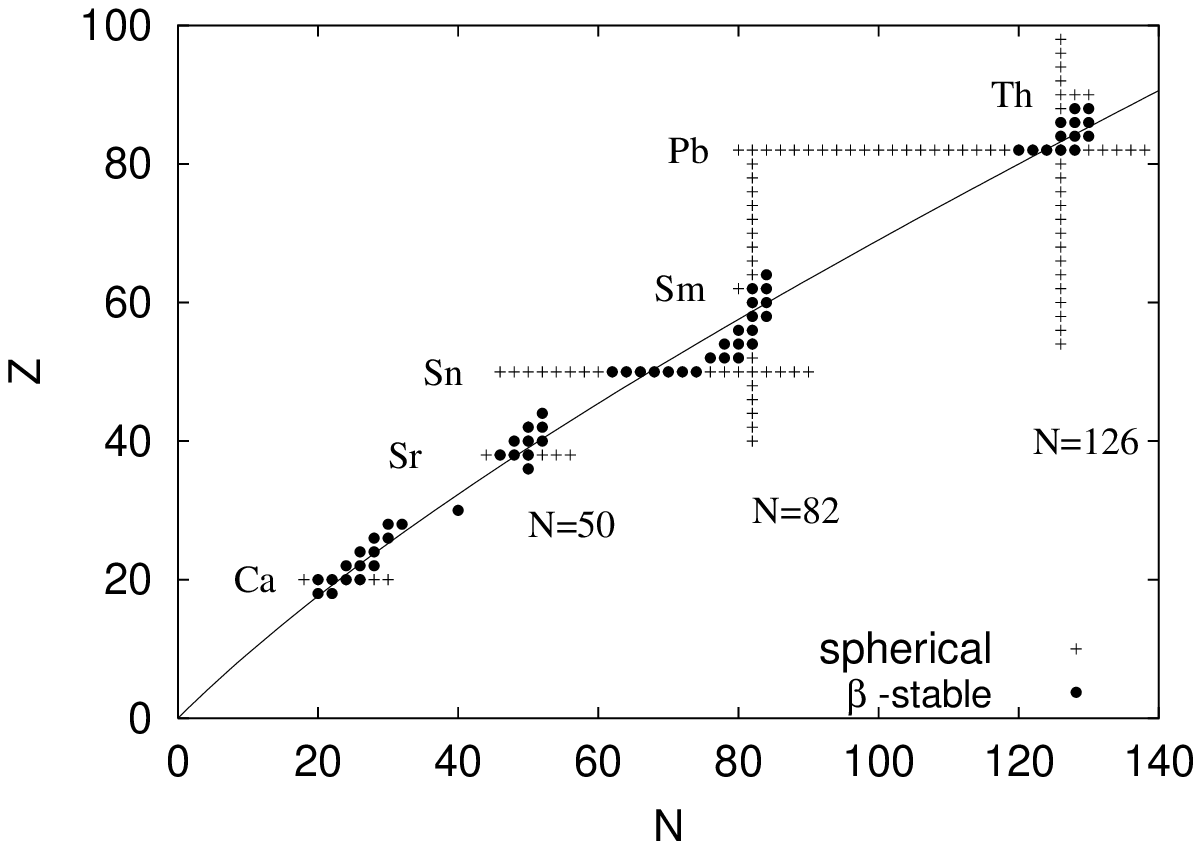, width=16.5cm, angle=00}

  \end{center}

  \caption{($N, Z$) plot of the spherical nuclei analysed in the present work.
The circles indicate the $\beta$-stable nuclei and
the crosses  a few representative isotope and isotone chains.}
\label{fig1} \end{figure}

\begin{figure}

  \begin{center}

  \leavevmode

\epsfig{file=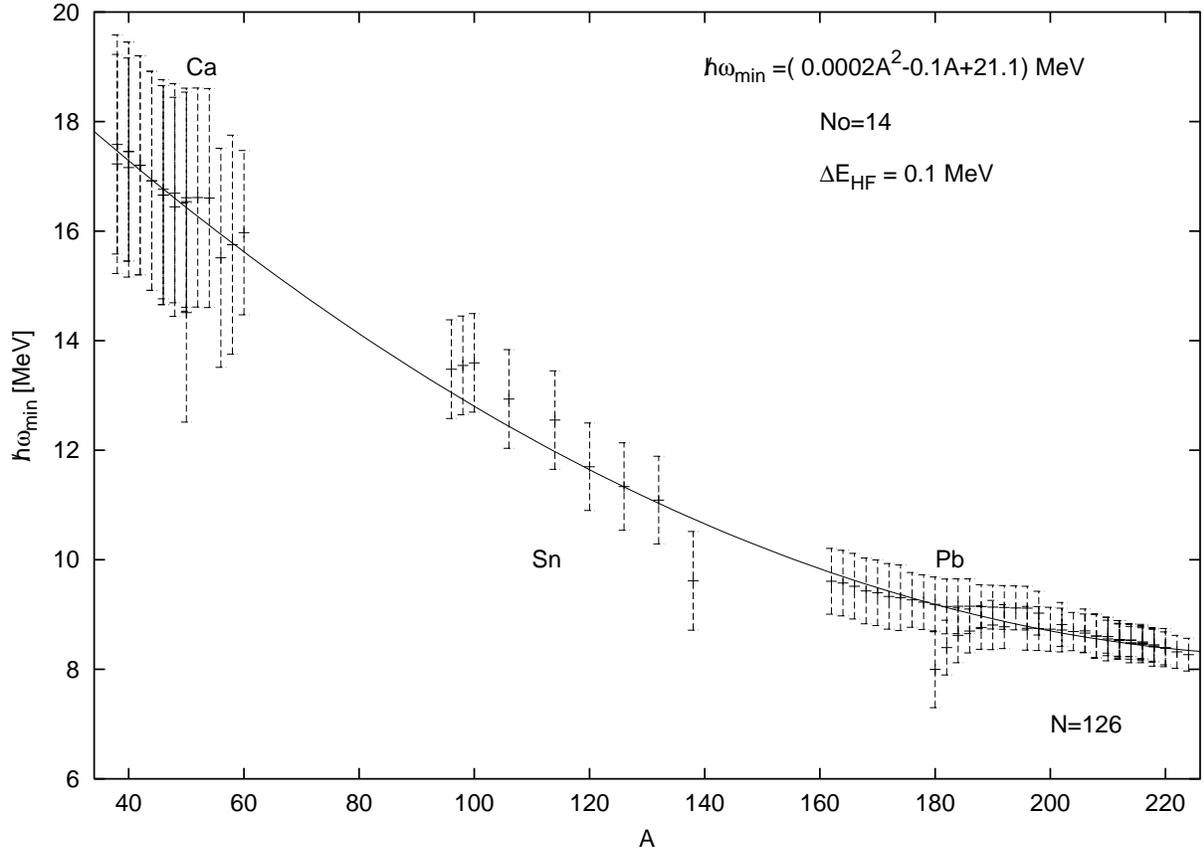, width=16.5cm, angle=00}

  \end{center}

  \caption{The dependence on $A$ of the parameter $\hbar\omega_{min}$
  of the $N_o$=14 HO bases employed in Ca, Sn and Pb isotopes.
The error bars indicate ranges of $\hbar\omega$ for which
the variation of the HF energy does not exceed 0.1 {\rm MeV}.}
  \label{fig2}
\end{figure}

\begin{figure}

  \begin{center}

  \leavevmode

\epsfig{file=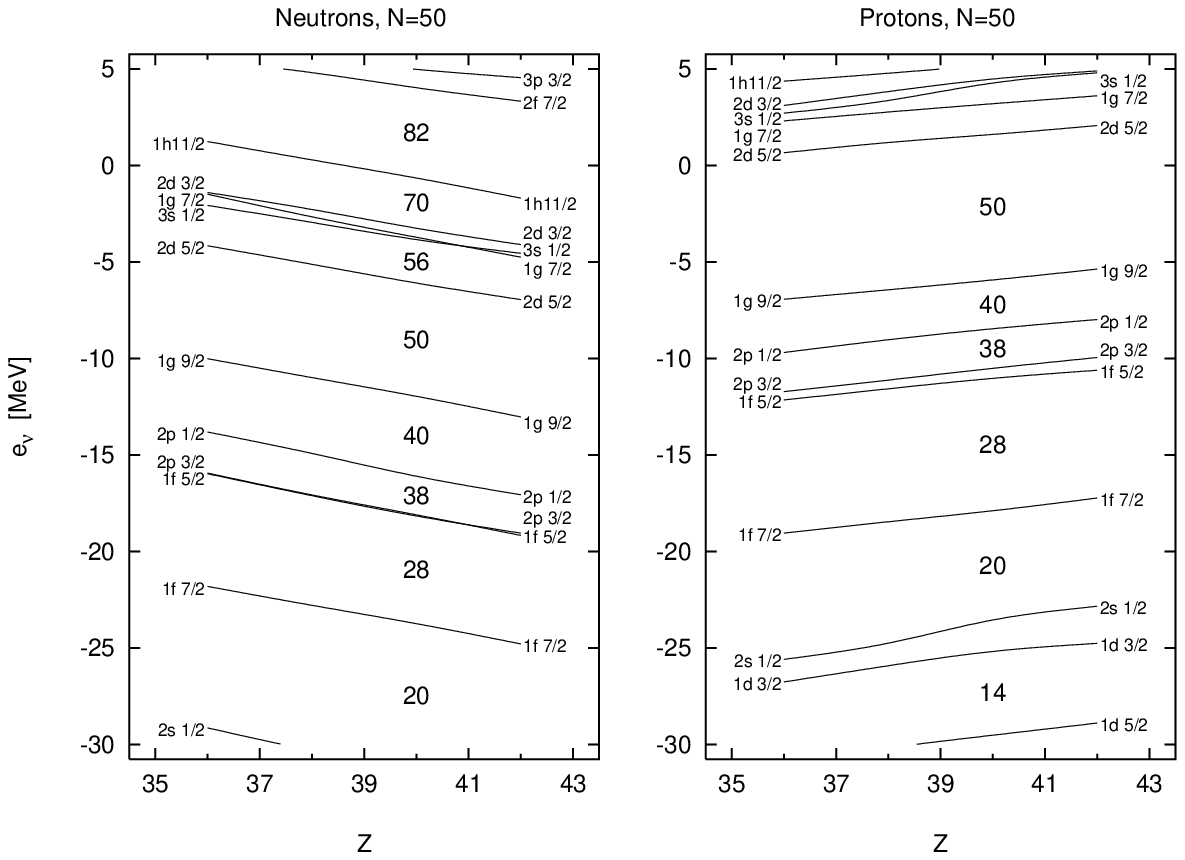, width=16.5cm, angle=00}

  \end{center}

  \caption{The neutron (left) and proton (right) single-particle levels
  in $N$=50 isotones versus the proton number $Z$.}
  \label{fig3a}
\end{figure}

\begin{figure}

  \begin{center}

  \leavevmode

\epsfig{file=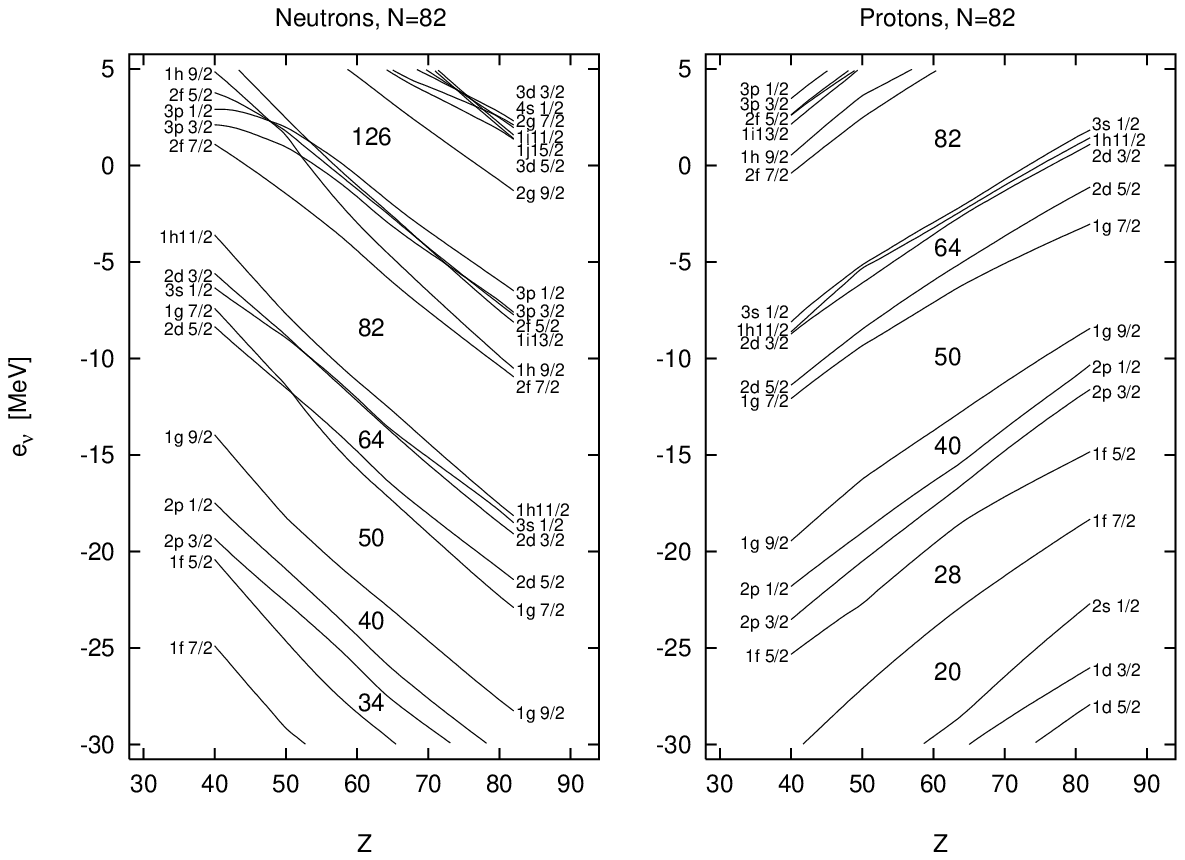, width=16.5cm, angle=00}

  \end{center}

  \caption{The neutron (left) and proton (right) single-particle levels
  in $N$=82 isotones versus the proton number $Z$.}
  \label{fig3b}
\end{figure}

\begin{figure}

  \begin{center}

  \leavevmode

\epsfig{file=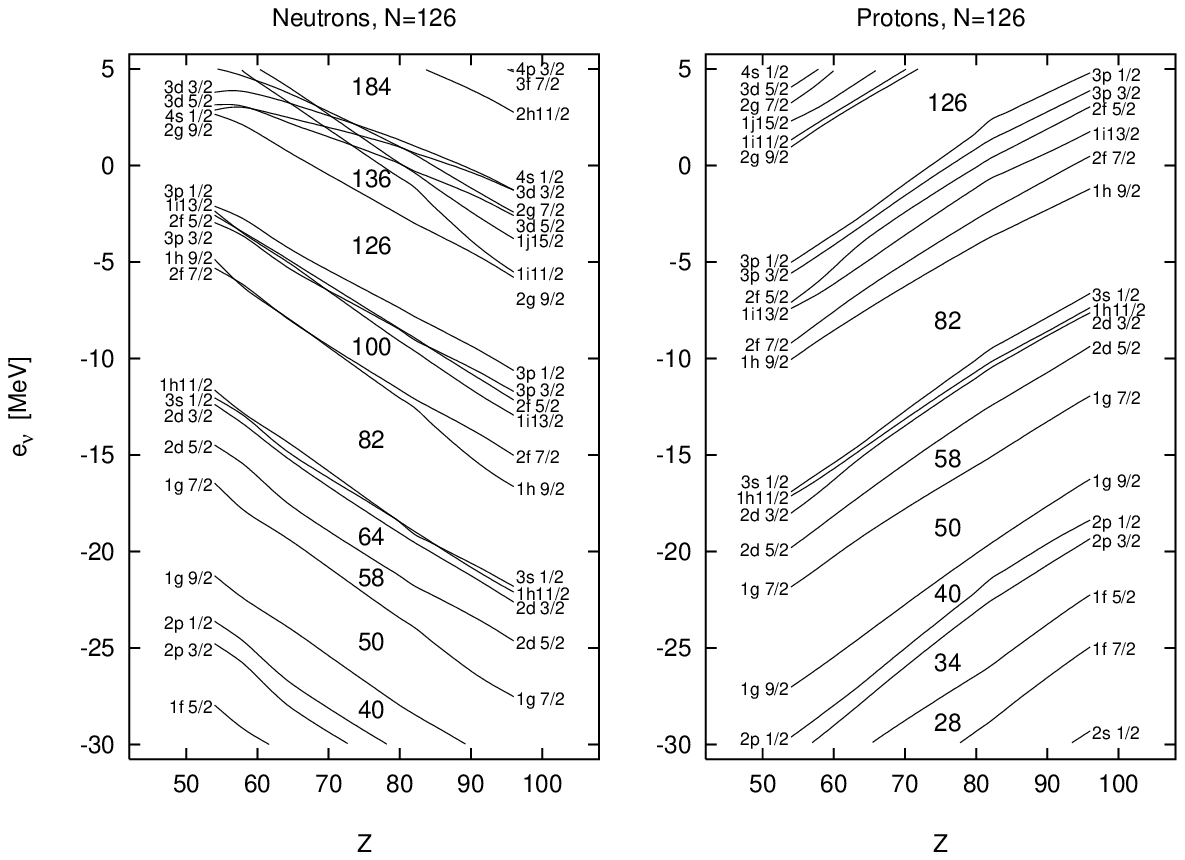, width=16.5cm, angle=00}

  \end{center}

  \caption{The neutron (left) and proton (right) single-particle levels
  in $N$=126 isotones versus the proton number $Z$.}
  \label{fig3c}
\end{figure}

\begin{figure}

  \begin{center}

  \leavevmode

\epsfig{file=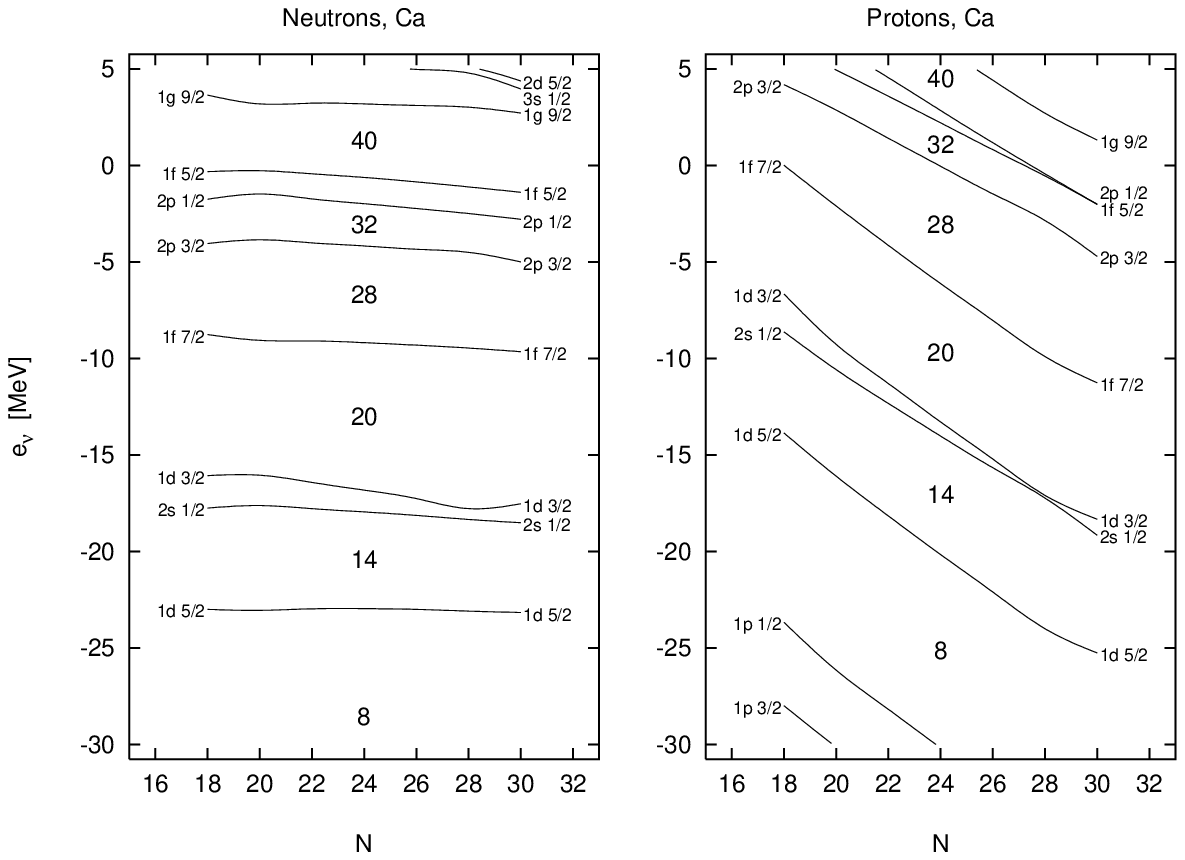, width=16.5cm, angle=00}

  \end{center}

  \caption{The neutron (left) and proton (right) single-particle levels
  in Ca  isotopes versus the neutron number $N$.}
  \label{fig4a}
\end{figure}

\begin{figure}

  \begin{center}

  \leavevmode

\epsfig{file=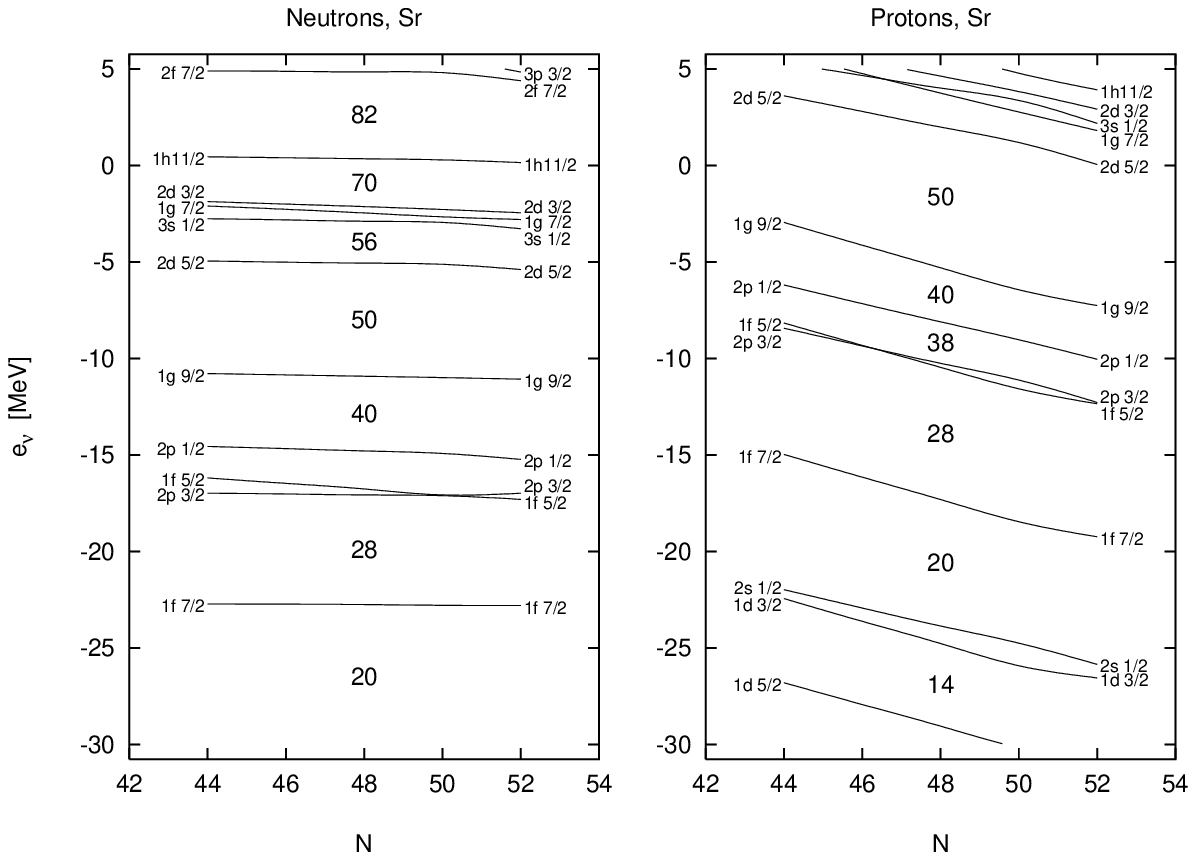, width=16.5cm, angle=00}

  \end{center}

  \caption{The neutron (left) and proton (right) single-particle levels
  in Sr  isotopes versus the neutron number $N$.}
  \label{fig4b}
\end{figure}

\begin{figure}

  \begin{center}

  \leavevmode

\epsfig{file=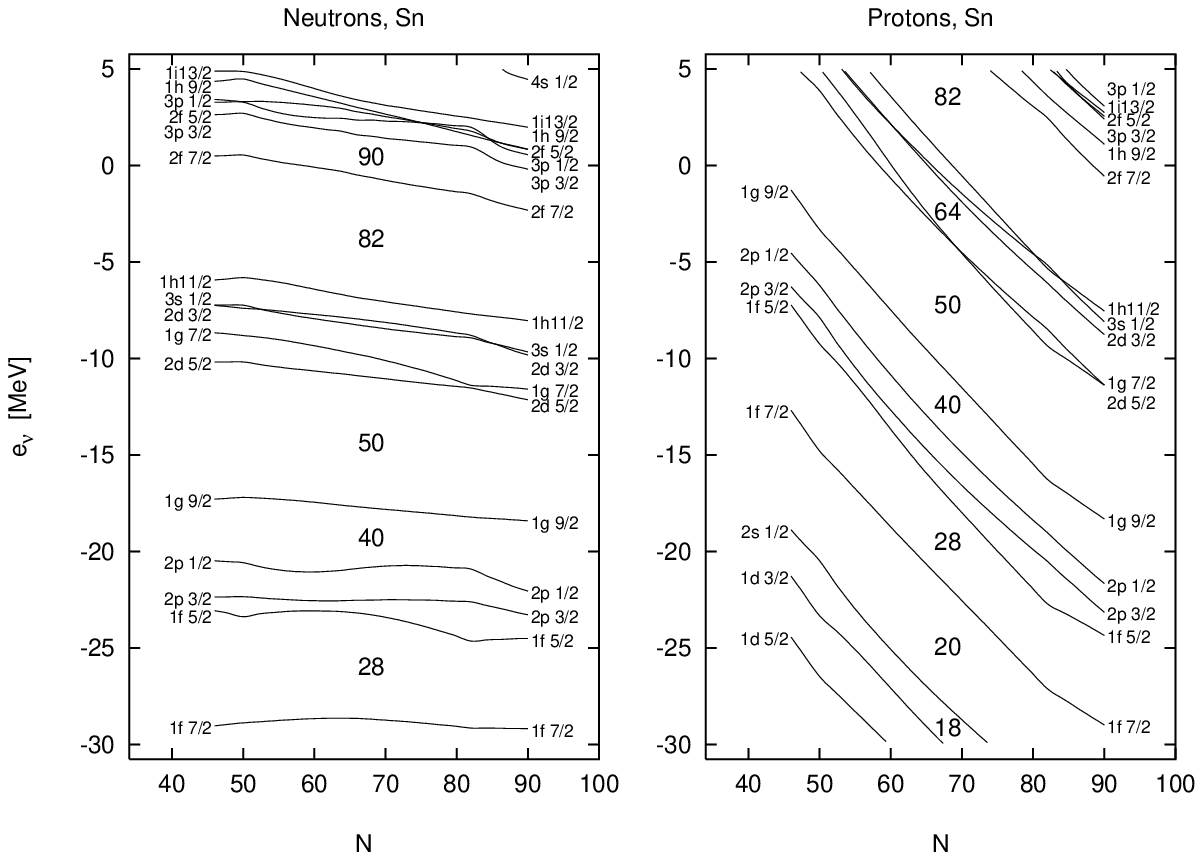, width=16.5cm, angle=00}

  \end{center}

  \caption{The neutron (left) and proton (right) single-particle levels
  in Sn  isotopes versus the neutron number $N$.}
  \label{fig4c}
\end{figure}

\begin{figure}

  \begin{center}

  \leavevmode

\epsfig{file=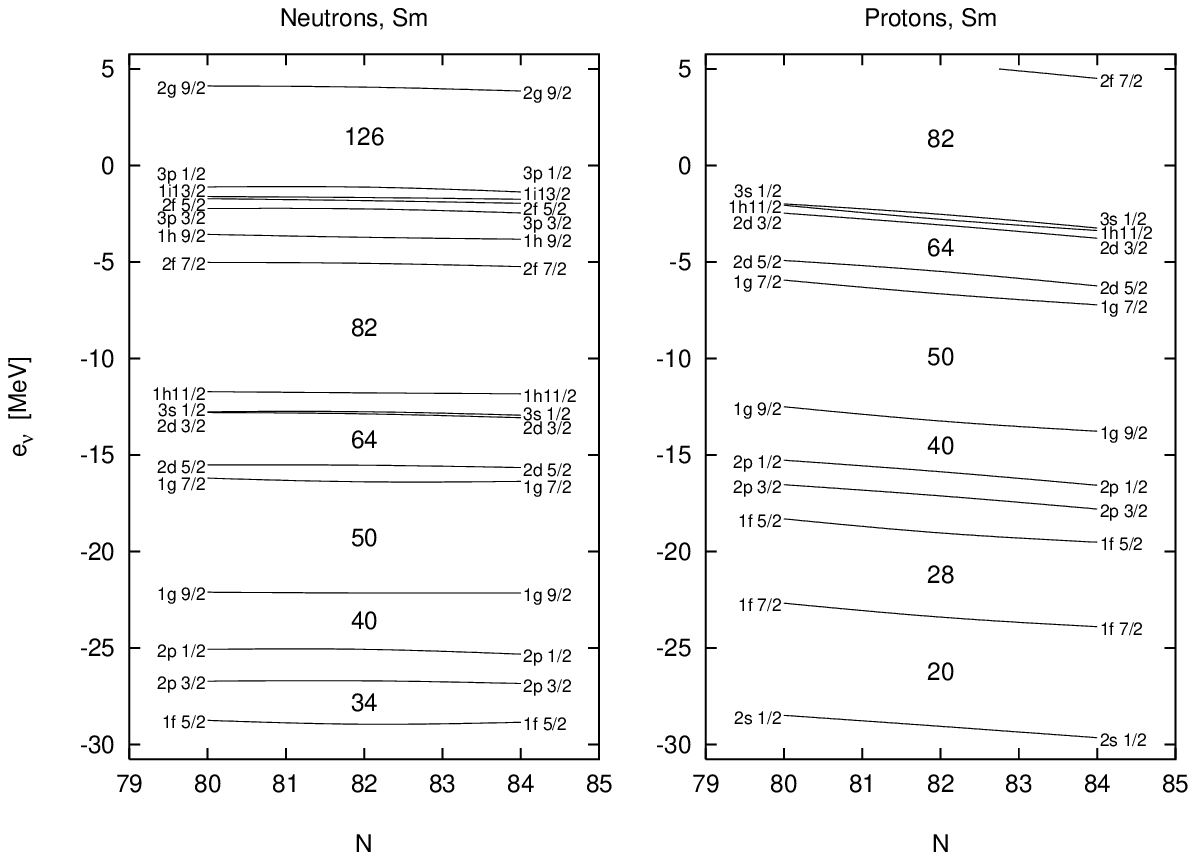, width=16.5cm, angle=00}

  \end{center}

  \caption{The neutron (left) and proton (right) single-particle levels
  in Sm  isotopes versus the neutron number $N$.}
  \label{fig4d}
\end{figure}

\begin{figure}

  \begin{center}

  \leavevmode

\epsfig{file=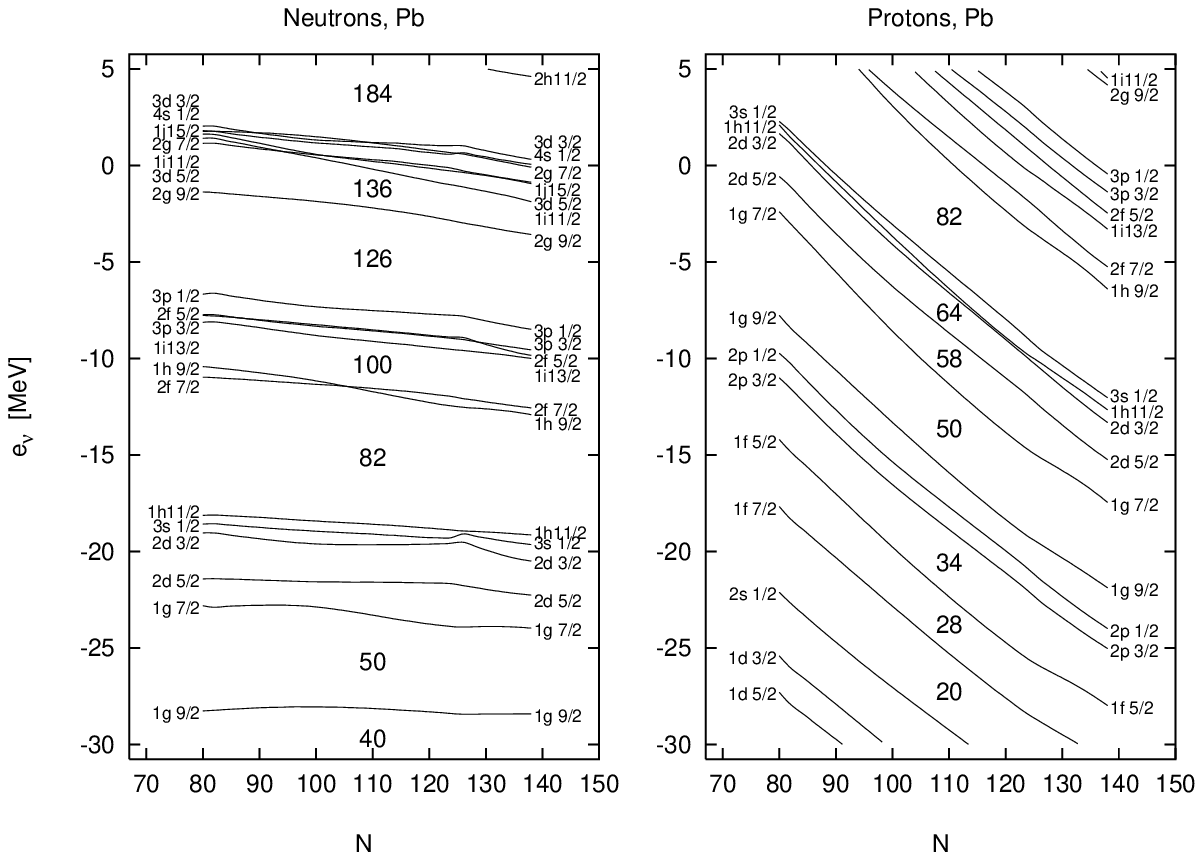, width=16.5cm, angle=00}

  \end{center}

  \caption{The neutron (left) and proton (right) single-particle levels
  in Pb  isotopes versus the neutron number $N$.}
  \label{fig4e}
\end{figure}

\begin{figure}

  \begin{center}

  \leavevmode

\epsfig{file=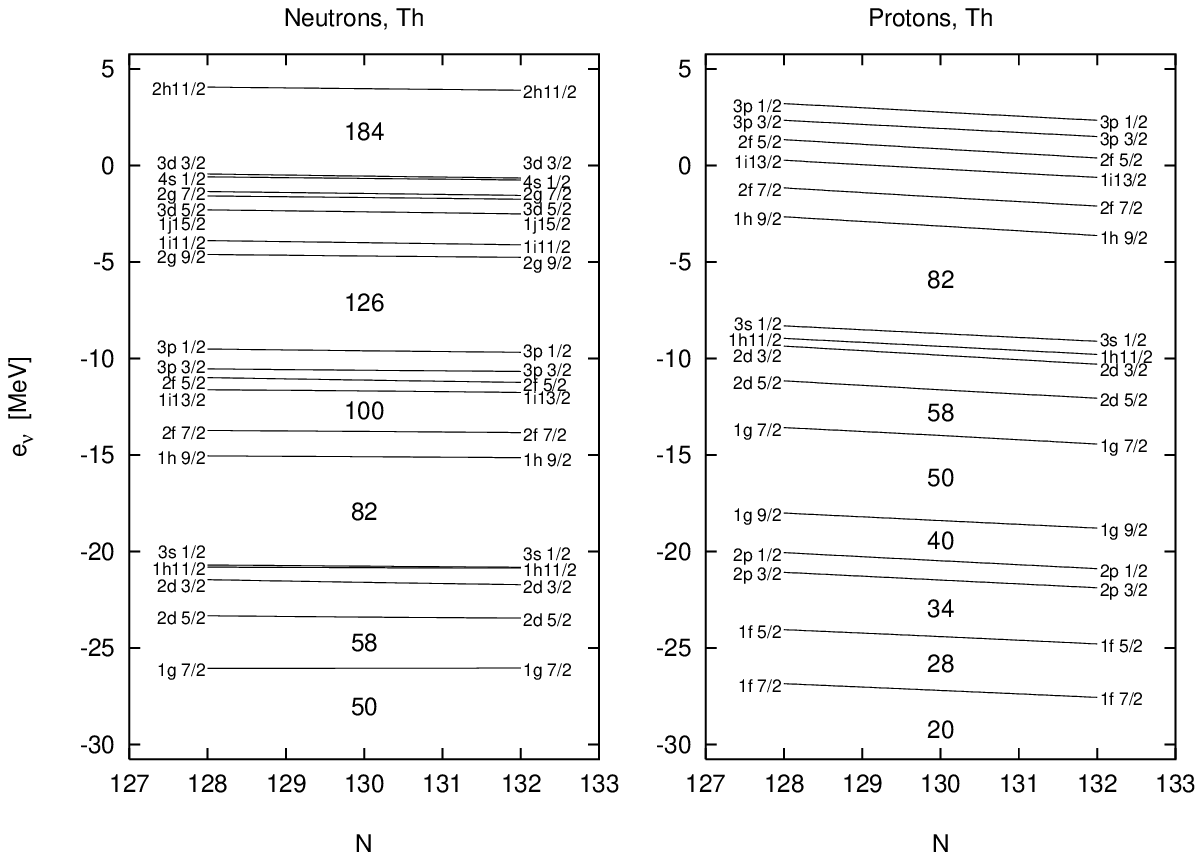, width=16.5cm, angle=00}

  \end{center}

  \caption{The neutron (left) and proton (right) single-particle levels
  in Th  isotopes versus the neutron number $N$.}
  \label{fig4f}
\end{figure}

\begin{figure}

  \begin{center}

  \leavevmode

\epsfig{file=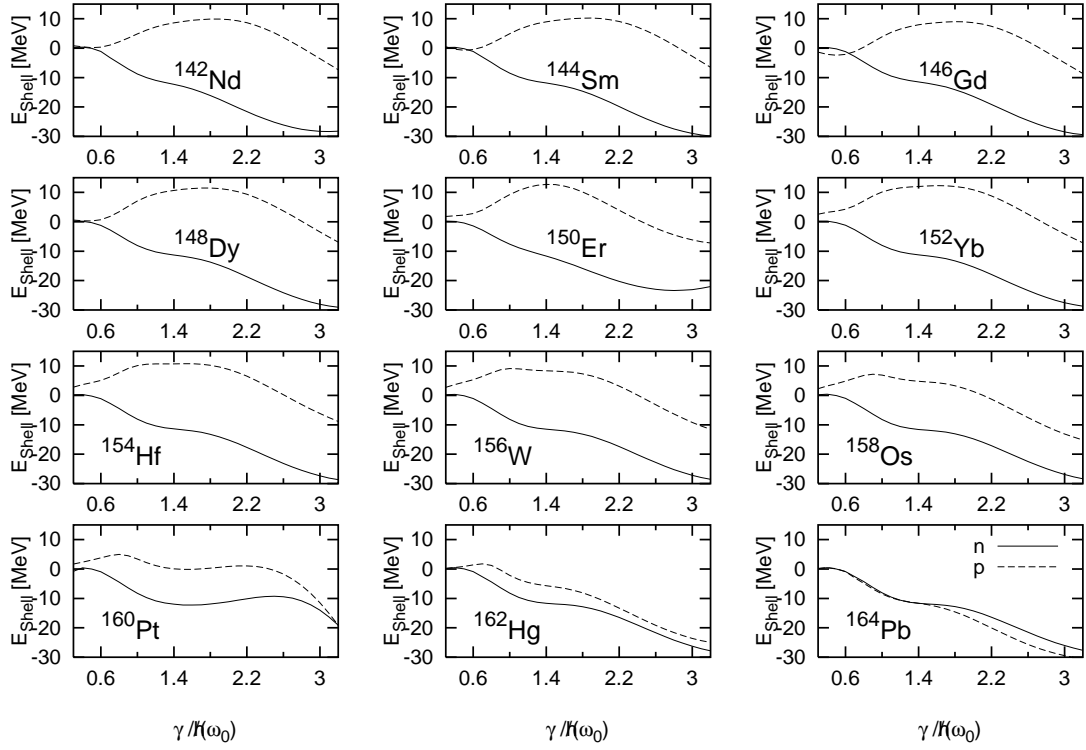, width=16.5cm, angle=00}

  \end{center}

  \caption{Proton (dashed lines) and neutron (solid lines)
  shell corrections of the Nd up to Pd $N$=82 isotones
  as functions of $\gamma / (\hbar\omega_{0})$
  with $\hbar\omega_{0} =  40 A^{-\frac{1}{3}} {\rm MeV}$.}
  \label{fig5}
\end{figure}

\begin{figure}

  \begin{center}

  \leavevmode

\epsfig{file=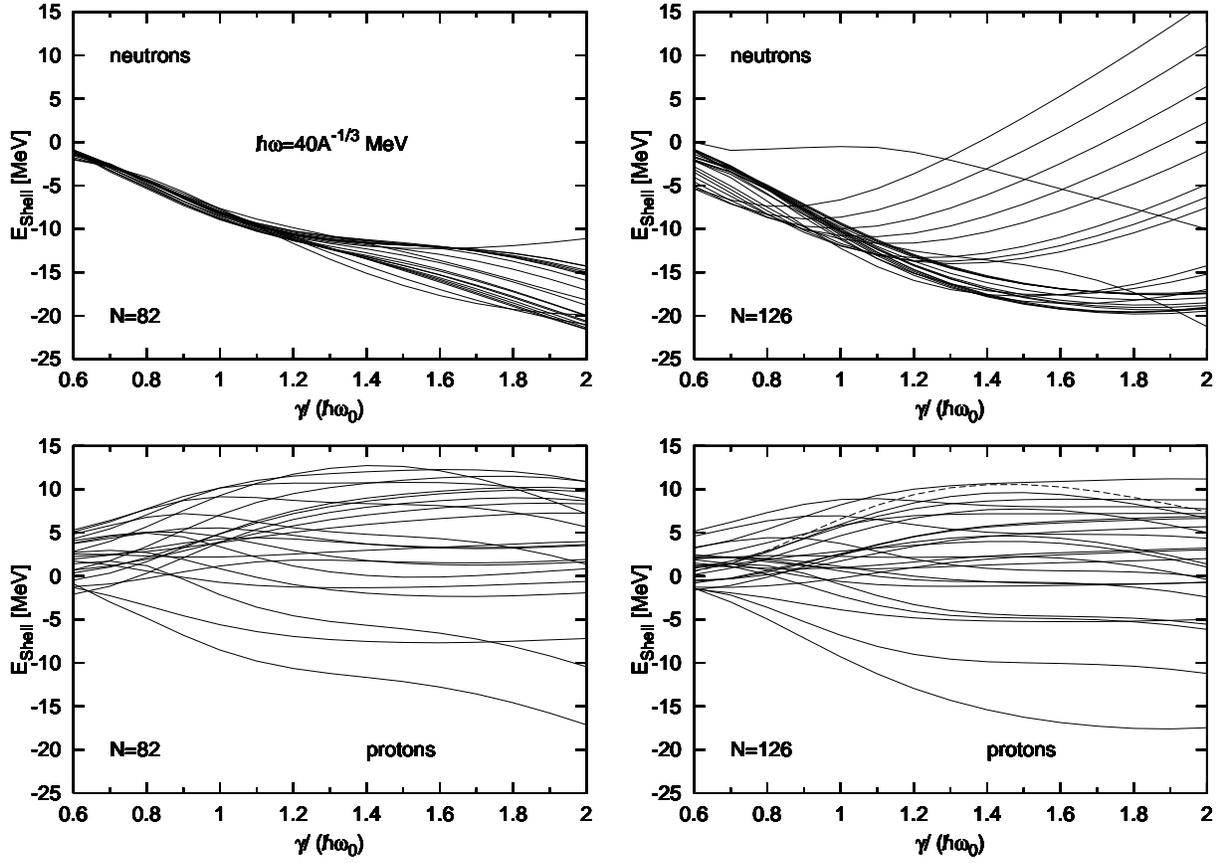, width=16.5cm, angle=00}

  \end{center}

  \caption{Shell corrections for neutrons (upper row) and protons (lower
          row) of $N$=82 (left) and $N$=126 (right) isotones versus
   $\gamma/(\hbar\omega_{0})$  with $\hbar\omega_{0} =
  40 A^{-\frac{1}{3}} {\rm MeV}$.}
  \label{fig6}
\end{figure}

In order to find a plateau in the variations of the shell correction with
respect to the width $\gamma$ appearing in eq. (\ref{rho}), $\gamma$ has
been varied, taking as a scale the average shell spacing
$\hbar\omega_{0} = 40 A^{-\frac{1}{3}} {\rm MeV}$
\cite{nil}.
In fact Strutinsky calculations generally obtain a plateau in the shell
correction energy of both protons and neutrons for $\gamma$ in the vicinity
of the traditional value $\gamma = 1.2 \hbar\omega_{0}$
\cite{nil}.
Fig.~\ref{fig5} and \ref{fig6} give examples of the variations
with $\gamma / (\hbar\omega_{0})$ of the shell corrections obtained
in the present work.
Similar behaviours are found in all the nuclei displayed in the ($N$,$Z$) plot of
Fig.~\ref{fig1}.
In Fig.~\ref{fig5},
the neutron (solid lines) and  proton (dashed lines) shell corrections
of twelve $N$=82 isotones ranging from Nd to Pb are plotted
as functions of $\gamma / (\hbar\omega_{0})$.
In Fig.~\ref{fig6}, the different curves represent the shell corrections
for neutrons (top) and protons (bottom)
calculated in  $N$=82 (left) and $N$=126 (right) isotones.
One can see that, except for a few cases, as the
proton shell corrections in $^{196}$Yb and $^{198}$Dy, reasonable plateaus
are always found  around the value $\gamma =1.2\,\hbar\omega_{0}$.
The results shown in the next figures are those obtained with the latter
value of the averaging width $\gamma$.

\begin{figure}

  \begin{center}

  \leavevmode

\epsfig{file=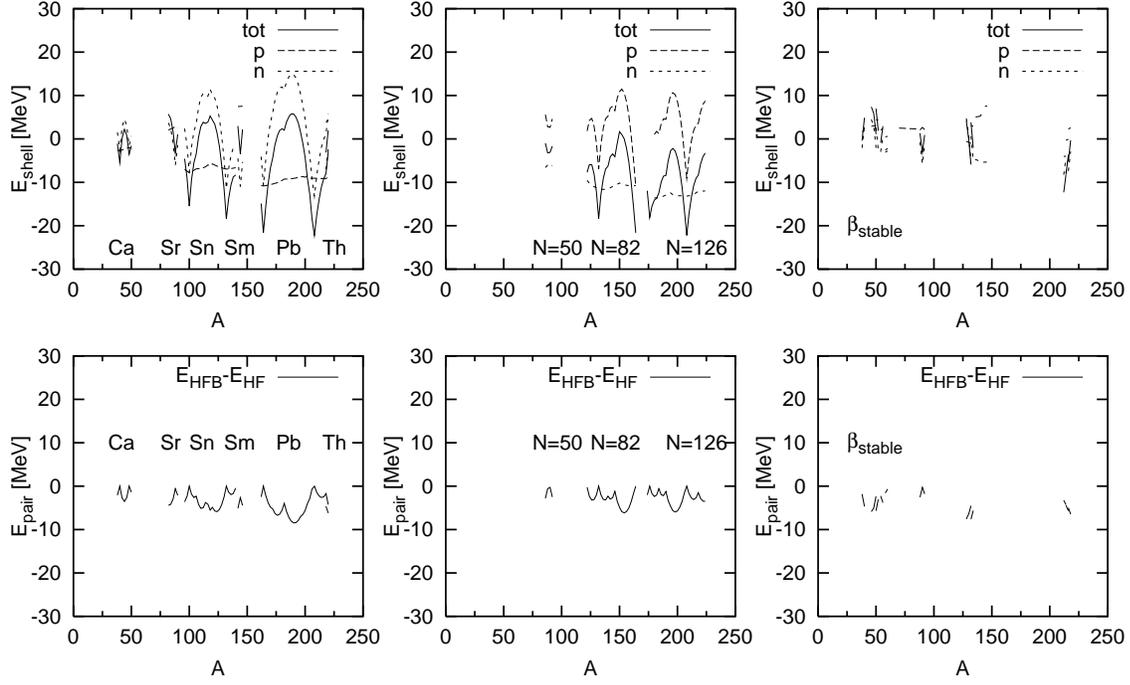, width=16.5cm, angle=00}

  \end{center}

  \caption{The upper row shows the total (solid lines), and neutron and
          proton (dashed lines) shell corrections
          of Ca up to Th isotopes (left),
   $N$=50,  $N$=82 and  $N$=126 isotones (center) and
   $\beta$-stable nuclei (right) versus the  mass number $A$.
   The lower row displays the pairing corrections
  calculated as the differences $ E_{\rm pair}=E_{HFB} - E_{HF}$
    for the same nuclei.}
  \label{fig7}
\end{figure}

In the upper part of Fig.~\ref{fig7} the  neutron  and  proton (dashed
lines), and total (solid lines) shell corrections are drawn as functions of
$A$ for the Ca  up to Th isotopes (left),   $N$=50, 82 and 126 isotones (center)
and $\beta$-stable nuclei (right) included in this study.
The lower part of this figure displays the corresponding pairing corrections
calculated from eqs. (\ref{pair}) .
One observes that the variations of the shell corrections nicely reproduce
the expected pattern, with
minima located at magic proton and neutron numbers.
Neutron shell corrections are seen to be almost independent of the
nucleus neutron number $N$,
as they should, and  proton shell corrections are  found practically independent
of $Z$.

\begin{figure}

  \begin{center}

  \leavevmode

\epsfig{file=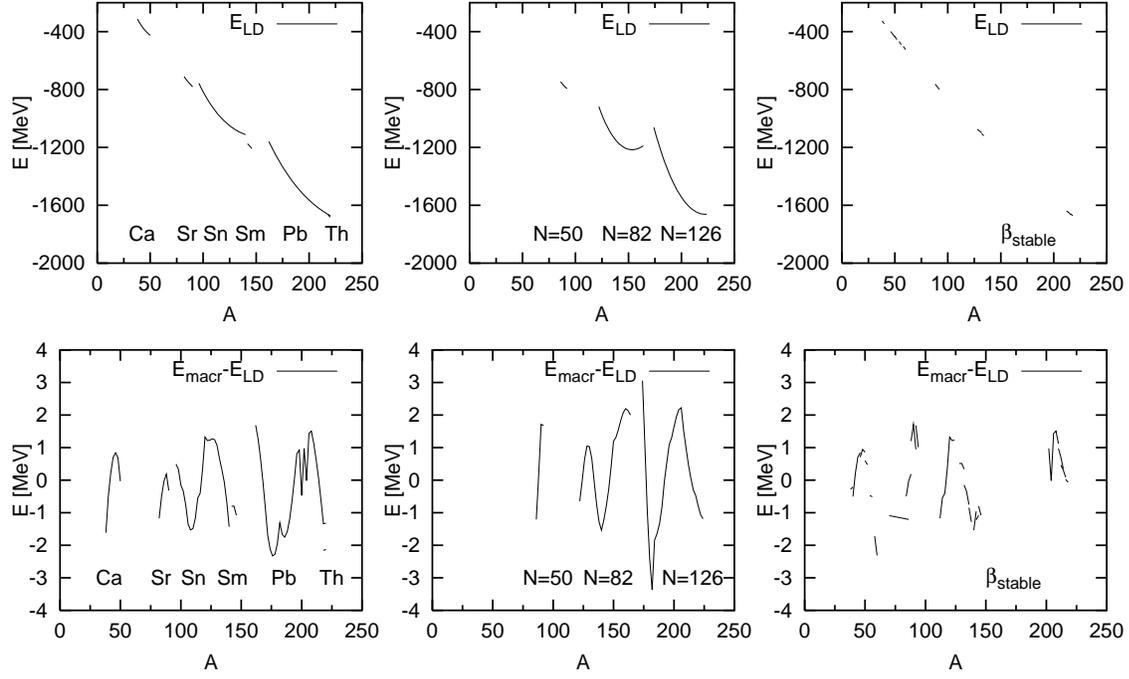, width=16.5cm, angle=00}

  \end{center}

  \caption{The fitted liquid drop energy $E_{\rm LD}$
  given by eq.
  (\ref{eld}) (solid lines) for constant $Z$ (left),
constant $N$  (center), and $\beta$-stable (right) nuclei versus their mass
number $A$.
The differences $E_{\rm macr}-E_{\rm LD}$ are shown in the three
lower plots for the same nuclei.}
  \label{fig8}
\end{figure}

The macroscopic energies (\ref{macr}) obtained by subtracting
the proton and neutron shell corrections from the  Hartree-Fock energies
$E_{HF}$  for all the nuclei displayed in Fig.~\ref{fig1}
are represented by lines in the three upper plots of  Fig.~\ref{fig8}.
A fit of these macroscopic energies using a liquid drop formula of the
form proposed by  Myers and \'Swi\c atecki \cite{mye} yields the following
expression:
\begin{equation}
 E_{\rm LD} = [15.65(1-1.92I^2) A -18.92(1-2.1I^2) A^{2/3}
  -0.73{ Z^2\over A^{1/3}}+1.99{ Z^2\over A}]\ {\rm MeV} \,.
\label{eld}
\end{equation}
with $I=(N-Z)/A$.
The numerical parameters in this formula are in good agreement
with those of  Myers and \'Swi\c atecki (M-S) \cite{mye},
except for the coefficient in the last term: 1.99 instead of 1.21.
This term represents the correction to the Coulomb energy of a charge
liquid drop that accounts for the diffuseness of the proton density
distribution. The correction appears to be larger in the case of formula
(\ref{eld}) --~the difference with the equivalent M-S term reaches $\simeq$
30 MeV in
$^{252}$Fm~--, which compensates for the fact that the
Coulomb energy derived from the self-consistent results is larger than
the M-S one. Let us also note that the asymmetry coefficients
in the volume and surface terms of (\ref{eld}) --~the coefficients of $I^2$~--
are not equal, contrary to what is assumed in the M-S formula.
On the other hand, the parameters appearing in (\ref{eld}) are closer to
those of the most recent best empirical fits
-~where the whole set of presently known nuclear masses and fission barriers
is used~- than to those of the various formulae derived from liquid drop
models \cite{pad}.

The fitted macroscopic energy (\ref{eld}) and the calculated values
$E_{\rm macr}$ are completely superimposed in the three upper plots
of Fig.~\ref{fig8}.
In order to better appreciate the quality of the fit given by eq. (\ref{eld}),
the differences $E_{\rm macr} - E_{\rm LD}$ are displayed
in the lower part of the figure, using an enlarged energy scale.
One observes that these differences do not exceed 3 MeV,
which represents less than 0.5 \% of the total $E_{HF}$ energy.

\section{Radii}
\label{Sect6}

In order to derive  systematics for the radii of all the nuclei
displayed in Fig.~\ref{fig1}, eq.
(\ref{ldr}) has been used
to extract from the different
-- neutron, proton, mass and charge -- Hartree-Fock-Bogoliubov r.m.s. radii, corresponding isospin-dependent radius constant
$r^n_0$, $r^p_0$ , $r^{\rm tot}_0$ and $r^{\rm ch}_0$.
Namely, each kind of microscopic r.m.s. radius is multiplied by
$\sqrt{\frac{5}{3}}A^{-\frac{1}{3}}$.
A fit of the obtained values with formula (\ref{rphen})
yielded the following parametrizations:
\begin{equation}
 r^n_0 = 1.17 \left(1 + 0.12 I + 3.29/A\right)\ {\rm fm}\,,
\label{r0n}
\end{equation}
\begin{equation}
 r^p_0 = 1.21 \left(1 - 0.14 I + 1.83/A\right)\ {\rm fm}\,,
\label{r0p}
\end{equation}
\begin{equation}
 r^{\rm tot}_0 = 1.19 \left(1 +0.03I + 2.70/A\right) \ {\rm fm}\,.
\label{r0tot}
\end{equation}
\begin{equation}
r^{\rm ch}_0 = 1.22 \left(1 - 0.15I + 2.32/A\right) \ {\rm fm}\,,
\label{r0ch}
\end{equation}
It is interesting to observe that the leading coefficient in (\ref{r0tot})
is almost equal to the usual value $r_0$= 1.2 fm, although the $I$ and $A$
dependence included in the next terms may lead to significant deviations
in the case of light or exotic nuclei.

The above parametrizations appear very close to those derived  in \cite{pom}
by taking into account the Ca, Sr, Sn, Sm and Pb isotopes only, and to
those obtained in the analysis of the r.m.s. radii calculated
in the framework of the relativistic mean field theory for a set of nuclei
similar to the one envisaged here \cite{war}.

Using formula (\ref{ror}), the ratios of
proton to neutron radii could be fitted with the parametrization:
\begin{equation}
  {r_{p}\over r_{n}} = 1.04 \left(1 - 0.27I  - 1.12/A\right) \,.
\label{rporn}
\end{equation}
This parametrization again appears very close to
those derived in the studies of Refs. \cite{pom,kleb,war} mentioned above.

Let us point out that the parametrization given by formula (\ref{rporn}),
which has been established for the spherical nuclei
shown in Fig.~\ref{fig1}, has been found later on to reproduce the ratios of
the HFB proton to neutron radii of all nuclei between the proton and
neutron drip lines, spherical or deformed,
in the range $20<A<318$. This latter result follows from large scale
HFB calculations with axial symmetry presently under way with the Gogny force.

\begin{figure}

  \begin{center}

  \leavevmode

\epsfig{file=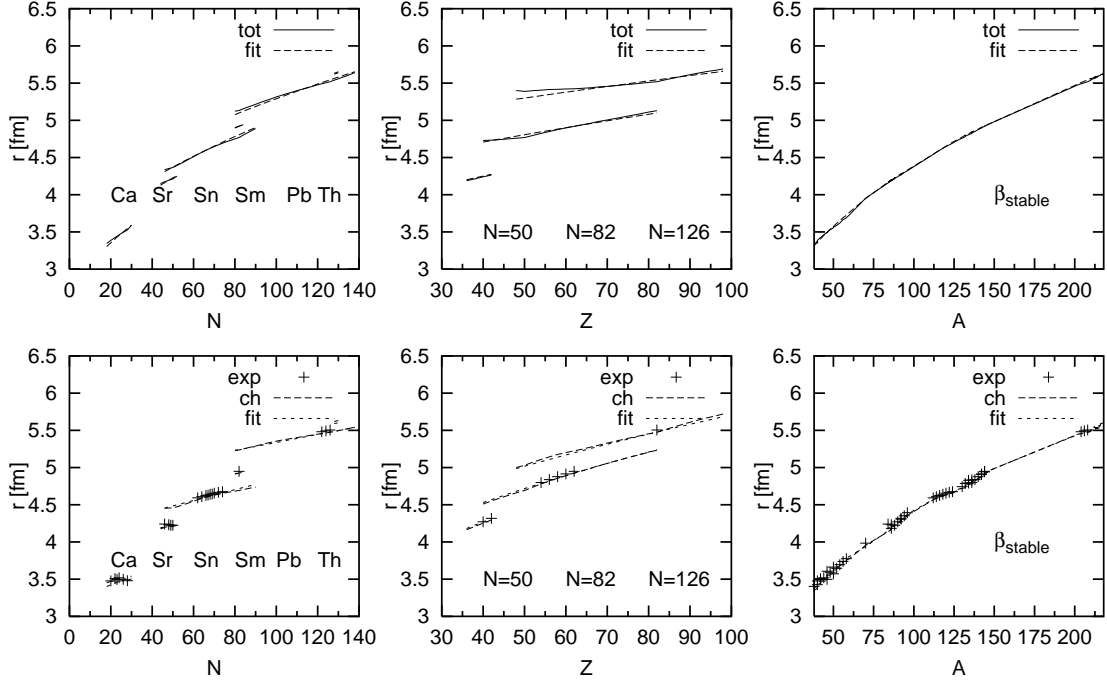, width=16.5cm, angle=00}

  \end{center}

  \caption{Upper row: total HFB r.m.s. radii (solid lines), and their
fits with formula (\ref{r0tot}) (dashed lines) in constant $Z$ (left),
constant $N$ (center)  and $\beta$-stable (right) nuclei.
Lower row: HFB charge radii (dashed lines) compared with
experimental data~\protect\cite{fri} (crosses) and with the fit given
by formula  (\ref{r0ch}).}
  \label{fig9}
\end{figure}

\begin{figure}

  \begin{center}

  \leavevmode

\epsfig{file=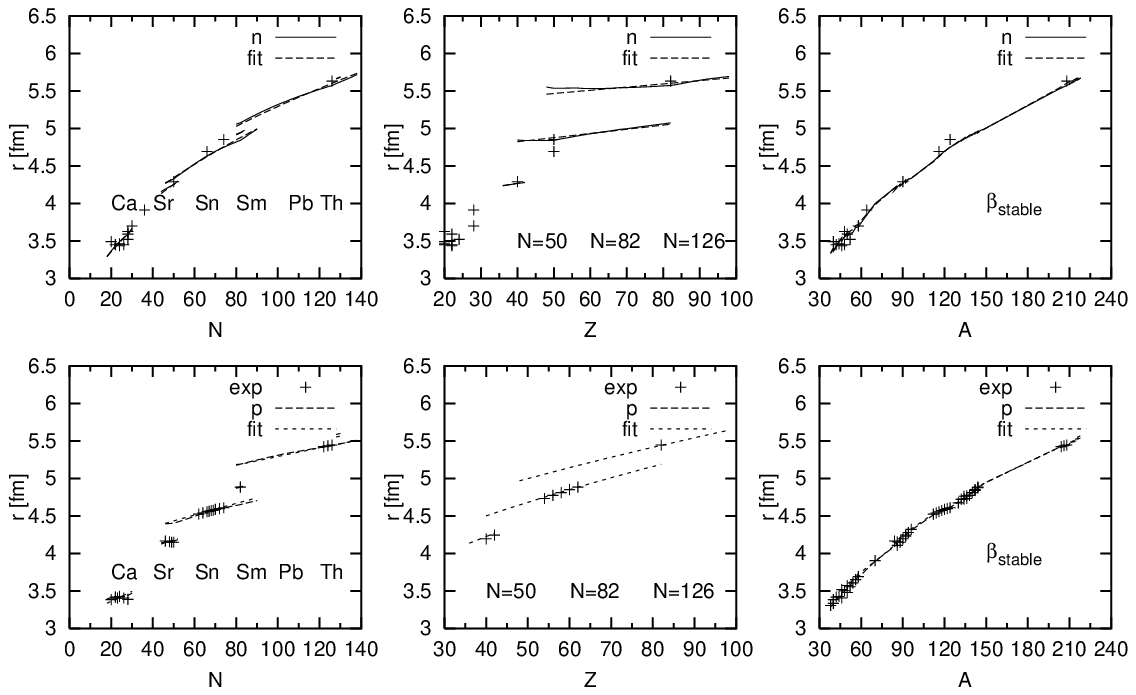, width=16.5cm, angle=00}

\end{center}

  \caption{Neutron (upper row, solid lines) and proton (lower row, dashed
          lines) HFB r.m.s. radii of constant $Z$ (left), constant  $N$
  (center), and $\beta$-stable (left)
          nuclei compared with experimental data \protect\cite{fri,bat}
   (crosses) and with formulas (\ref{r0n})-(\ref{r0p}) (dashed lines).}
  \label{fig10}
\end{figure}

The results obtained in the present work
are summarized in Figs.~\ref{fig9} and \ref{fig10}.
The upper part of Fig.~\ref{fig9} displays the mass HFB radii
(solid lines) at constant $Z$ (left),  constant $N$ (center)
and for $\beta$-stable (right) nuclei, together with
the fits given by eq. (\ref{r0tot}).
The lower part of the figure shows in a similar way the results obtained for
charged radii. The fits are given by eq. (\ref{r0ch}), and experimental data,
taken from Ref.~\cite{fri}, have been indicated by crosses.
Calculated charge radii appear in good agreement with
experimental ones, except in a few cases.
One must stress in this respect that the charge radii  obtained
in the HFB approach do not include the influence
of the long range correlations associated with
collective oscillations of the mean-field, such as RPA ground state
correlations or shape coexistence effects occurring in soft nuclei.
As well known, these correlations beyond the mean field
may lead to a significant increase of HFB r.m.s. radii.

Fig.~\ref{fig10} is the equivalent of Fig.~\ref{fig9} for neutron
 (upper ) and proton (lower part) radii.
Experimental data on proton radii (crosses) have been taken
from Ref.~\cite{fri,bat} and the fits are given by formulae
(\ref{r0n}) and (\ref{r0p}).

Finally, as already pointed out in Section~\ref{Sect2}, the
ratios $r_{p}/r_{n}$ are deformation-independent quantities that can be used
to calculate the neutron radii of nuclei whose charged
radii are known (see eq. (\ref{rn})). From formula (\ref{rporn}), one gets
the following expression for neutron radii as functions of $A$,
$I=(N-Z)/A$ and $\langle r_{ch}^{2}\rangle$, the mean square charge radius:
\begin{equation}
r_{n}=\frac{\sqrt{\langle r_{ch}^{2}\rangle  - 0.64}}{1.04(1-0.27I -1.12/A)}\
      {\rm fm}\,.
\label{ern}
\end{equation}

\section{Conclusions}
\label{Sect7}

  In this work, the results of self-consistent HFB calculations performed
with the Gogny effective interaction for several isotopic and isotonic chains
of spherical nuclei have been analysed and compared with those given by
phenomenological expressions and by the relativistic mean field approach.
The more important conclusions that can be drawn from this analysis are the
following:
\begin{enumerate}

\item The average binding energy of nuclei calculated by applying the
Strutinsky smoothing procedure to the single-particle level scheme
obtained from the HFB self-consistent mean field, eq. (\ref{eld}), is in
excellent agreement with the liquid-drop formula of Myers and \'Swi\c atecki
(M-S) \cite{mye}.
The main difference occurs in the term correcting the charged liquid drop
Coulomb energy for the diffuseness of the proton density.
Also, the asymmetry coefficients in the volume and surface terms are
found unequal, contrary to the M-S formula,
and the reduced radius parameter $r_{0}$ is slightly smaller
($r_{0}$ = 1.19 fm instead of $r_{0}$ = 1.205 fm).
When all contributions are added, the differences between eq. (\ref{eld})
and the M-S formula compensate, yielding nuclear binding
energies in agreement within 3 MeV.

\item The r.m.s. of the neutron and charge
distributions
agree in a quite satisfactory way with experimental data. From the set of nuclei
studied in this work, systematics of the neutron, proton, charge and mass
r.m.s. radii are obtained  for spherical nuclei
in the form of parametrizations depending on
the mass number $A$ and asymmetry $I=(N-Z)/A$
(see eqs. (\ref{r0n})-(\ref{r0ch})).
These parametrizations are of the same form as those first proposed in
Ref.\cite{nerl}. The values of the parameters found in the present work are
consistent with those derived from experimental data \cite{nerl} and from
the microscopic calculations performed in the framework of the
relativistic mean field approach \cite{rin}.

\item Systematics of the ratio $r_{p}/r_{n}$ of the proton r.m.s.   to 
neutron r.m.s. radius  are of special interest because
they are, to a very good approximation, independent of the nuclear
deformation and, in addition, they can be used to determine the value
of the neutron radius of any nucleus, either spherical or deformed, when its
charge radius is known (see eq. (\ref{rn})).
The parametrization of $r_p / r_n$ obtained from the set of nuclei
considered in the present study (eq. (\ref{rporn}))
appears in excellent agreement with previous ones based either on
experimental data  \cite{nerl} or on the relativistic mean field approach
\cite{rin}. This parametrization can therefore be considered as robust
enough to be used for predictions concerning nuclei for which experimental
data  not yet available, such as nuclei close to the
neutron drip line or superheavy elements.
\end{enumerate}

\acknowledgments
 
 We  wish to express our thanks to Krzysztof Pomorski 
 for the helpful discussions. The careful reading of the manuscript
 done by Jony Bartel is also acknowledged.
 B.N.P. is very grateful for the nice hospitality extended to her by the
 Service de Physique Nucl\'eaire of the  
 Centre d'Etudes de Bruy\`eres-le-Ch\^atel.



%


%

\newpage

\newpage

\newpage

\newpage

\newpage

\newpage

\newpage

\newpage

\newpage

\newpage

\newpage

\newpage

\newpage

\newpage

\newpage

\newpage

\newpage

\newpage

\begin{thebibliography}{10}

\bibitem{dech} J. Decharg\'e and D. Gogny, Phys. Rev. {\bf C21}, 1568 (1980).

\bibitem{ber} J. F. Berger, M. Girod, and D. Gogny, Computer Physics Communication,
              {\bf 63}, 365 (1991), Nucl. Phys. {\bf A428}, 23c (1984).

\bibitem{pom} M. Kleban and B. Nerlo-Pomorska, Annales UMCS {\bf AAA LV/LVI},
       1 (2000/2001).

\bibitem{kleb} M. Kleban, B. Nerlo-Pomorska, K. Pomorski, J.F. Berger, and
        J. Decharg\'e, Acta Phys. Polon. B {\bf 32}, 1119 (2001).

\bibitem{mye} W. D. Myers and W. J. \'Swi\c atecki, Nucl. Phys.
        {\bf 81}, 1 (1966),
        Ark. Phys.{\bf 36}, 343 (1967).
\bibitem{rin} K. Pomorski, P. Ring, G. A. Lalazissis, A. Baran, Z. \L{}ojewski,
        B. Nerlo-Pomorska, and M. Warda, Nucl. Phys. {\bf A624}, 349 (1997).

\bibitem{strut} V. M. Strutinsky, Yad. Fiz. {\bf 36}, 614 (1966).

\bibitem{nerl} B. Nerlo-Pomorska and K. Pomorski, Z. Phys. {\bf A348},
        169 (1994).

\bibitem{war} M. Warda, B. Nerlo-Pomorska, and K. Pomorski, Nucl. Phys.
       {\bf A635}, 484 (1998).

\bibitem{mol}P. M\"oller, J. P. Nix, and K. Z. Kratz, At. Data Nucl. Data
        Tables {\bf 66}, 131 (1997).

\bibitem{nil} S. G. Nilsson, C. F. Tsang, A. Sobiczewski, Z. Szyma\'nski, S.
        Wycech, C. Gustafson, I. L. Lamm, P. M\"oller, and B. Nilsson,
        Nucl. Phys. {\bf A131}, 1 (1969).

\bibitem{pad} K. Pomorski, J. Dudek, private communication.
 
\bibitem{fri} G. Fricke, C. Bernardt, K. Heiling, L. A. Schaller, 
        L. S. Schellenberg, E.B. Shera, C. W. De Jager,
	 At. Data Nucl. Data Tabl. {\bf 60},
              177 (1995).

\bibitem{bat} C. J. Batty, E. Friedman, H. J. Gils, and H. Rebel,
        Adv.  Nucl. Phys. {\bf 19}, 1 (1989).

\end{thebibliography}
\end{document}